# A Study of IEEE 802.15.4 Security Framework for Wireless Body Area Networks

Shahnaz Saleem, Sana Ullah, and Kyung Sup Kwak

Graduate School of Information & Communication Engineering, Inha University,
253 Yonghyun-dong, Nam-gu, Incheon 402-751, Korea

Email: {roshnee13, sanajcs}@hotmail.com, kskwak@inha.ac.kr

**Abstract:** A Wireless Body Area Network (WBAN) is a collection of low-power and lightweight wireless sensor nodes that are used to monitor the human body functions and the surrounding environment. It supports a number of innovative and interesting applications, including ubiquitous healthcare and Consumer Electronics (CE) applications. Since WBAN nodes are used to collect sensitive (life-critical) information and may operate in hostile environments, they require strict security mechanisms to prevent malicious interaction with the system. In this paper, we first highlight major security requirements and Denial of Service (DoS) attacks in WBAN at Physical, Medium Access Control (MAC), Network, and Transport layers. Then we discuss the IEEE 802.15.4 security framework and identify the security vulnerabilities and major attacks in the context of WBAN. Different types of attacks on the Contention Access Period (CAP) and Contention Free Period (CFP) parts of the superframe are analyzed and discussed. It is observed that a smart attacker can successfully corrupt an increasing number of GTS slots in the CFP period and can considerably affect the Quality of Service (QoS) in WBAN (since most of the data is carried in CFP period). As we increase the number of smart attackers the corrupted GTS slots are eventually increased, which prevents the legitimate nodes to utilize the bandwidth efficiently. This means that the direct adaptation of IEEE 802.15.4 security framework for WBAN is not totally secure for certain WBAN applications. New solutions are required to integrate high level security in WBAN.

## 1. Introduction

A Wireless Body Area Network (WBAN) allows the integration of intelligent, miniaturized, low-power sensor nodes in, on, or around a human body to monitor body functions and the surrounding environment. It has great potential to revolutionize the future of healthcare technology and has attracted a number of researchers both from the academia and industry in the past few years. WBANs support a wide range of medical and Consumer Electronics (CE) applications. For example, WBANs provide remote health monitoring of patients for a long period of time without any restriction on his/her normal activities [1,2]. Different nodes such as Electrocardiogram (ECG), Electromyography (EMG), and Electroencephalography (EEG) are deployed on the human body to collect the physiological parameters and forward them to a remote medical server for further analysis as given in Figure 1. Generally WBAN consists of in-body and on-body area networks. An in-body area network allows communication between invasive/implanted devices and a base station. An on-body area network, on the other hand, allows communication between non-invasive/wearable devices and a base station.

**Figure 1.** WBAN architecture for medical applications.

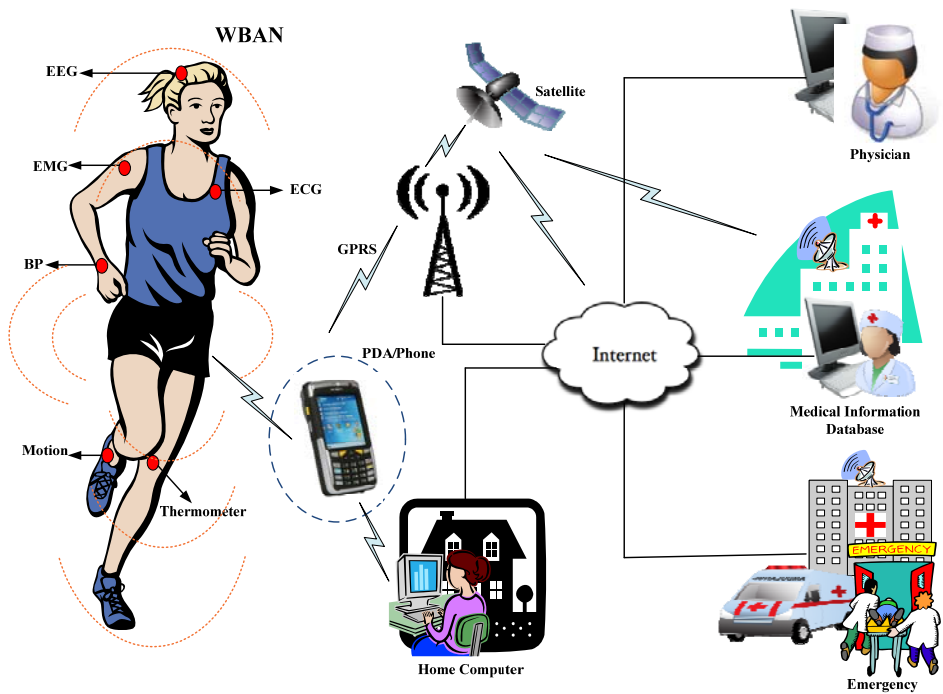

The consideration of WBANs for medical and non-medical applications must satisfy stringent security and privacy requirements. These requirements are based on different applications ranging from medical (heart monitoring) to non-medical (listening to MP4) applications [3]. In case of medical applications, the security threats may lead a patient to a dangerous condition, and sometimes to death. Thus, a strict and scalable security mechanism is required to prevent malicious interaction with WBAN. A secure WBAN should include confidentiality and privacy, integrity and authentication, key establishment and trust set-up, secure group management and data aggregation. However, the integration of a high-level security mechanism in a low-power and resource-constrained sensor node increases the computational, communication and management costs. In WBANs, both security and system performance are equally important, and thus, designing a low-power and secure WBAN system is a fundamental challenge to the designers. In this paper, we present a brief discussion on the major security requirements and threats in WBANs at the Physical, Medium Access Control (MAC), Network, and Transport layers. We analyze the performance of IEEE 802.15.4 [4,5] security framework for WBAN using extensive simulations. Different types of attack on IEEE 802.15.4 superframe are considered in the simulations. The results are presented for smart, random, and weak attackers in terms of probability of failed Guaranteed Time Slots (GTS) requests (due to backoff manipulation attacks) in the Contention Access Period (CAP) period, number of corrupted slots in the Contention Free Period (CFP) period, and decrease in bandwidth utilization. It is concluded that smart attackers can successfully disrupt the entire communication channel in the network.

The rest of the paper is categorized into six sections. Section 2 and Section 3 outline the major security issues and threats in WBAN. Section 4 describes the IEEE 802.15.4 security framework for WBAN. In Section 5, we identify possible attacks on the IEEE 802.15.4 superframe structure. Section 6 presents simulation results. The final section concludes our work.

## 2. Security Issues and Requirements

A WBAN is a special type of network which shares some characteristics with traditional WSNs but differs in many others such as strict security and low-power consumption. It is mandatory to understand the type of WBAN applications before the integration of a suitable security mechanism. The correct understanding will lead us towards a strong security mechanism that will protect the system from possible threats. The key security requirements in WBANs are discussed below.

### *2.1. Data Confidentiality*

Like WSNs, Data confidentiality is considered to be the most important issue in WBANs. It is required to protect the data from disclosure. WBANs should not leak patient's vital information to external or neighbouring networks. In medical applications, the nodes collect and forward sensitive data to the coordinator. An adversary can eavesdrop on the communication, and can overhear the critical information. This eavesdropping may cause severe damage to the patient since the adversary can use the acquired data for many illegal purposes. The standard approach to protect the data secure is to encrypt it with a secure key that can only be decrypted by the intended receivers. The use of symmetric key encryption is the most reliable for WBANs since public-key cryptography is too costly for the energy-constraint sensor nodes.

## 2.2. Data Integrity

Keeping the data confidential does not protect it from external modifications. An adversary can always alter the data by adding some fragments or by manipulating the data within a packet. This packet can later be forwarded to the coordinator. Lack of data integrity mechanism is sometimes very dangerous especially in case of life-critical events (when emergency data is altered). Data loss can also occur due to bad communication environment.

## 2.3. Data Authentication

It confirms the identity of the original source node. Apart from modifying the data packets, the adversary can also change a packet stream by integrating fabricated packets. The coordinator must have the capability to verify the original source of data. Data authentication can be achieved using a Message Authentication Code (**MAC**) (to differentiate it from Medium Access Control (MAC), the Message Authentication Code (**MAC**) is represented by bold letters) that is generally computed from the shared secret key.

## 2.4. Data Freshness

The adversary may sometimes capture data in transit and replay them later using the old key in order to confuse the coordinator. Data freshness implies that the data is fresh and that no one can replay old messages. There are two types of data freshness: weak freshness, which guarantees partial data frames ordering but does not guarantee delay, and strong freshness, which guarantees data frames ordering as well as delay.

## 2.5. Secure Localization

Most WBAN applications require accurate estimation of the patient's location. Lack of smart tracking mechanisms allow an attacker to send incorrect reports about the patient's location either by reporting false signal strengths or by using replaying signals.

## 2.6. Availability

Availability implies efficient availability of patient's information to the physician. The adversary may target the availability of WBAN by capturing or disabling a particular node, which may sometimes result in loss of life. One of the best ways is to switch the operation of a node that has been attacked to another node in the network.

*2.7. Secure Management*

Secure management is required at the coordinator to provide key distribution to the nodes for encryption and decryption operation. In case of association and disassociation, the coordinator adds or removes the nodes in a secure manner.

## 3. Possible Security Threats and Attacks

A WBAN is vulnerable to a considerable number of key attacks. These attacks are conducted in different ways, *i.e.*, Denial of Service (DoS) attacks, privacy violation, and physical attacks. Due to restrictions on the power consumption of the sensor nodes, protection against these types of attacks is a challenging task. A powerful sensor can easily jam a sensor node and can prevent it from collecting patient's data on regular basis.

Attacks on WBAN can be classified into three main categories [6]: (a) attacks on secrecy and authentication, where an adversary performs eavesdropping, packet replay attacks, or spoofing of packets, (b) attacks on service integrity, where the network is forced to accept false information [7], and (c) attacks on network availability (DoS attacks), where the attacker tries to reduce the network's capacity. In the following section, we briefly present most important DoS attacks at physical, data link, network, and transport layers. A brief summary of these attacks is given in Table 1 [8].

**Table 1.** WBAN OSI layers and DoS attacks/denfeses.

| Layers | DoS Attacks | Defenses |
|---|---|---|
| Physical | Jamming | Spread-spectrum, priority messages, lower duty cycle, region mapping, mode change |
| | Tampering | Tamper-proof, hiding |
| Link | Collision | Error correcting code |
| | Unfairness | Small frames |
| | Exhaustion | Rate limitation |
| Network | Neglect and greed | Redundancy, probing |
| | Homing | Encryption |
| | Misdirection | Egress filtering, authorization monitoring |
| | Black holes | Authorization, monitoring, redundancy |
| Transport | Flooding | Client Puzzles |
| | De-synchronization | Authentication |

*3.1. Physical Layer Attacks*

Some of the main responsibilities of physical layer include frequency selection and generation, signal detection, modulation, and encryption [9]. Since the medium is radio-based, jamming the network is always possible. The most common attacks are jamming and tampering. Jamming refers to interference with the radio frequencies of the nodes. The jamming source can be powerful enough to disrupt the entire network. Tampering refers to the physical attacks on the sensor nodes [10]. However, nodes in WBAN are deployed in close proximity to the human body, and this reduces the chances of physical tampering.

### 3.2. Data Link Layer Attacks

This layer is responsible for multiplexing, frame detection, channel access, and reliability. Attacks on this layer include creating collision, unfairness in allocation, and resource exhaustion. Collision occurs when two or more nodes attempt to transmit at the same time. An adversary may strategically create extra collisions by sending repeated messages on the channel. Unfairness degrades the network performance by interrupting the MAC priority schemes. Exhaustion of battery resources may occur when a self-sacrificing node always keeps the channel busy.

### 3.3. Network Layer Attacks

The nodes in WBAN are not required to route the packets to other nodes. Routing is possible when multiple WBANs communicate with each other through their coordinators. Possible attacks include spoofing, selective forwarding, sybil, and hello flood. In spoofing, the attacker targets the routing information and alters it to disrupt the network. In selective forwarding, the attacker forwards selective messages and drops the others [11]. In sybil, the attacker represents more than one identity in the network [12]. The hello flood attacks are used to fool the network, *i.e.*, the sender is within the radio range of the receiver.

### 3.4. Transport Layer Attacks

The attacks on the transport layer are flooding and de-synchronisation. In flooding, the attacker repeatedly places requests for connection until the required resources are exhausted or reach a maximum limit. In de-synchronisation, the attacker forges messages between nodes causing them to request the transmission of missing frames.

## 4. IEEE 802.15.4 Security for WBAN

IEEE 802.15.4 is a low-power standard designed for low data rate applications. It offers three operational frequency bands: 868 MHz, 915 MHz, and 2.4 GHz bands. There are 27 sub-channels allocated in IEEE 802.15.4, *i.e.*, 16 sub-channels in 2.4 GHz band, 10 sub-channels in 915 MHz band and one sub-channel in the 868 MHz band. IEEE 802.15.4 MAC has two operational modes: a beacon-enabled mode and a non-beacon enabled mode. In the beacon-enabled mode, the network is controlled by a coordinator, which regularly transmits beacons for device synchronization and association control. The channel is bounded by a superframe structure as illustrated in Figure 2.

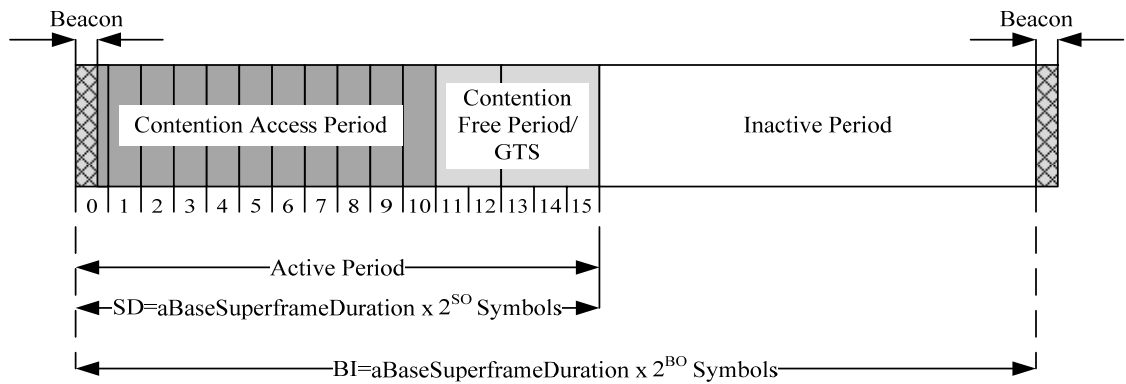
Figure 2. IEEE 802.15.4 superframe structure.

The superframe consists of both active and inactive periods. The active period contains three components: a beacon, a Contention Access Period (CAP), and a Contention Free Period (CFP). The coordinator interacts with nodes during the active period and sleeps during inactive period. There are maximum of seven GTS slots in the CFP period to support time critical traffic. In the beacon-enabled mode, a slotted CSMA/CA protocol is used in the CAP period. In the non-beacon enabled mode, the channel is accessed using unslotted CSMA/CA protocol.

The main security requirements presented in the IEEE 802.15.4 standard specification are access control, confidentiality, frame integrity, and sequential freshness. Access control ensures the protection of frames from unauthorized nodes. Confidentiality makes sure that only legitimate nodes share the secret information. Frame integrity protects the frames from manipulation by an adversary. Sequential freshness confirms the freshness of the frames.

The IEEE 802.15.4 security layer is handled at the MAC layer. The security requirements are specified at the application layer by tuning some control parameters. If no parameters are selected, no security mechanism is used. The specification defines four packet types: beacon, data, acknowledgement, and control packets. The beacon packets are used for synchronization and resource allocation. No security information can be included in the acknowledgement packets. In others, the information such as integrity protection and confidentiality protection can be integrated whenever required. The IEEE 802.15.4 specification has a choice of security suites that control different security levels. Each security suite has different security properties, protection levels, and frame formats. The IEEE 802.15.4 based security suites can be considered for WBAN with necessary modifications. Table 2 lists different security suites defined in the IEEE 802.15.4 standard [13]. They are broadly classified into null, encryption only (AES-CTR), authentication only (AES-CBC-**MAC**), and encryption and authentication (AES-CCM) suites. In AES-CTR, confidentiality protection is provided using Advance Encryption Standard (AES) block cipher [14] with counter mode. In AES-CBC-**MAC**, security including integrating protection is provided using CBC-**MAC** [15]. The AES-CCM provides high-level security that includes both data integrity and encryption. Details about these security suites are presented in the standard.

Table 2. Security modes in IEEE 802.15.4

| Name | Description | Access Control | Confidentiality | Frame Integrity | Sequential Freshness |
|---|---|---|---|---|---|
| Null | No security | | | | |
| AES-CTR | Encryption only, CTR Mode | X | X | | X |
| AES-CBC-**MAC**-128 | 128 bit **MAC** | X | | X | |
| AES-CBC-**MAC**-64 | 64 bit **MAC** | X | | X | |
| AES-CBC-**MAC**-32 | 32 bit **MAC** | X | | X | |
| AES-CCM-128 | Encryption & 128 bit **MAC** | X | X | X | X |
| AES-CCM-64 | Encryption & 64 bit **MAC** | X | X | X | X |
| AES-CCM-32 | Encryption & 32 bit **MAC** | X | X | X | X |

The IEEE 802.15.4 is considered very close to WBAN due to its quick implementation, reliable security mechanism, and support of low data rate applications with low cost of power consumption. A significant improvement has been seen in the IEEE 802.15.4 in terms of superframe variation (expanding the CFP period) and contention access mechanisms [16,17]. Since contention access mechanisms are not reliable for WBAN due to Clear Channel Assessment (CCA) and heavy collision problems, researchers have urged to shrink the CAP period in the IEEE 802.15.4 superframe and subsequently extend the CFP period [18]. The purpose was to carry loads of packets in the CFP part of the superframe. As discussed earlier, the IEEE 802.15.4 specification defines seven GTS slots for collision free transmission. A node interested to grab the slot tracks the beacon for resource allocation. The coordinator decides the assignment of the GTS slot. If needed, more than one GTS slot can be allocated to a node. Figure 3(a,b) shows the GTS allocation and deallocation process defined in the IEEE 802.15.4 specification.

Figure 3. (a) GTS allocation process, (b) GTS deallocation process.

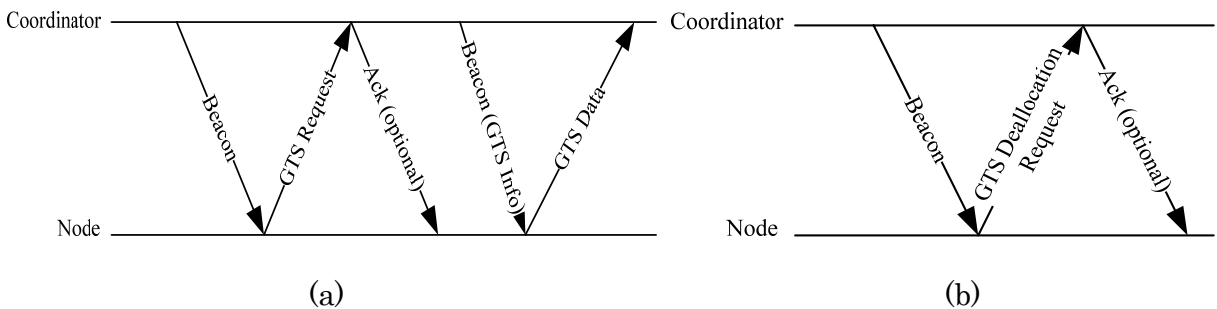

First, the nodes receive the beacons to identify the superframe boundaries. A GTS request is sent in the CAP part of the superframe to the coordinator. The request includes the required length and direction (uplink or downlink) of the GTS slot. The coordinator may send an acknowledgement packet to confirm the successful reception of the GTS request. If GTS slots are available, the coordinator assigns them to the nodes using the beacon frame. Once assigned, the data transmission takes place in the GTS slots of the following superframes.

The GTS allocation process may frequently occur in case of WBAN, where many nodes request the allocation of GTS slots. The main disadvantage of the IEEE 802.15.4 is the number of GTS slots is limited to seven. In WBAN, nodes generally require more GTS slots in the CFP period. This can be achieved by the varying the CFP duration according to the applications. No matter how many GTS slots are present in the CFP period, they have a vulnerable point that allows an attacker to disrupt the communication between nodes and the coordinator. Another problem is that the adversary may continuously select a small backoff window and may contend with the legitimate nodes (in the CAP period) in order to protect them from sending the GTS request packets. The following section briefly describes possible attacks on the CAP and CFP periods.

## 5. Attacks on the CAP and CFP Periods

Since most the traffic in WBAN is carried in the CFP period of the superframe, attacks on both CAP (this is used for resource allocation in CFP) and CFP periods can disrupt the entire communication between nodes and the coordinator. To attack the CAP period (also called backoff manipulation attack), a selfish node or an attacker attempts to select a small backoff window in order to keep the channel busy all the time. This attack prevents the legitimate nodes to send GTS slot requests to the coordinator as given in Figure 4(a). The backoff manipulation attack was first investigated for IEEE 802.11 networks in [19], where a selfish user implemented a whole range of strategies to maximize its access to the medium. Most of the challenging task is to detect backoff manipulation attacks [20,21]. Because the backoff counter is selected on random basis, it is very hard to identify the adversary who has deliberately chosen a small backoff window. A scheme to detect backoff manipulation attack is presented in [21], which works well for adversaries who are unaware of the detection scheme. But a smart adversary can efficiently maximize his throughput and can minimize the chances of his detection [22]. Another method of detecting these attacks is proposed in [23] where the receiver is used to assign backoff windows to the sender but the problem is that receiver cannot always be trusted. To attack the CFP period, an attacker carefully listens to the GTS allocation process and extracts the GTS slot information from the beacon [24] as given in Figure 4(b). The attacker first synchronizes itself to the network and receives periodic beacons. Assume that the legitimate node sends a GTS request to the coordinator. The attacker waits for the following beacon to extract the GTS slot information. Once the coordinator approves the GTS request, it integrates the slot information into the beacon frame. Both the legitimate node and the adversary receive the beacon. After obtaining the GTS slot information, the adversary can easily create interference in the GTS slot. Since the GTS slots are used to carry critical data (life-critical in case of WBANs [25]), interference in transmission affects the QoS requirements.

Figure 4. (a) Backoff manipulation attack on the CAP, (b) Attack on CFP period.

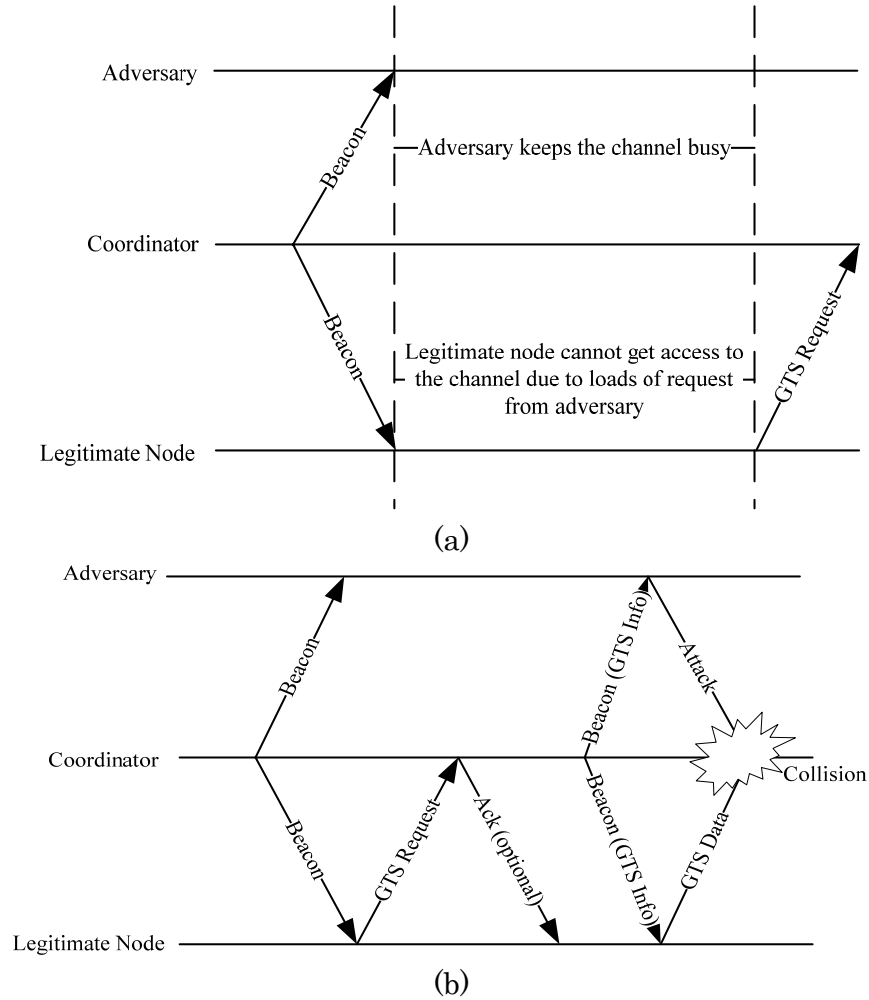

(a)

(b)

## 6. Evaluation and Results

We simulate a number of attacks on the CAP and CFP periods of the IEEE 802.15.4 superframe using the NS 2.31 simulator [26]. The simulation is based on the framework defined in [24]. We consider a network of ten legitimate nodes, which can be randomly attacked by five attackers. The attackers are categorized into smart, random, and weak attacks. Smart attackers aim at corrupting both the CAP and CFP periods. They corrupt the GTS slot with maximum duration. Random attackers aim at corrupting CFP period only with an average GTS slot duration. Weak attackers aim at corrupting GTS slots with minimum duration. The attacks are triggered at random basis in each simulation run and the results are analyzed in terms of probability of failed GTS requests (due to backoff manipulation attacks) in the CAP period, number of corrupted slots in the CFP period, and decrease in bandwidth utilization.

The smart attackers repeatedly attempt to access the channel in the CAP period, thus increasing the probability of failed GTS requests, as given in Figure 5. It can be seen that few smart attackers can disrupt the entire communication channel. Since the original data transmission in WBAN takes place in the CFP period, analysis of attacks on the CFP period is becoming increasingly important.

**Figure 5.** Probability of failed GTS requests.

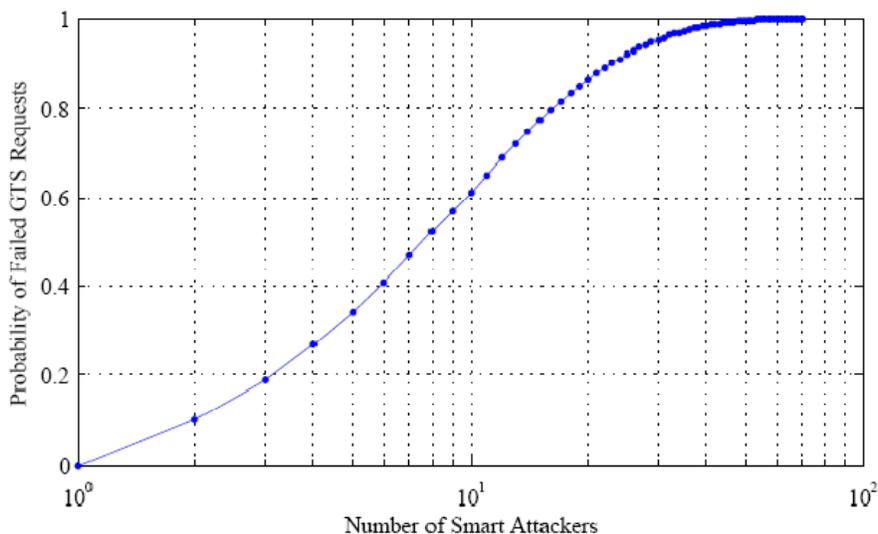

**Figure 6.** Total number of corrupted slots in the CFP.

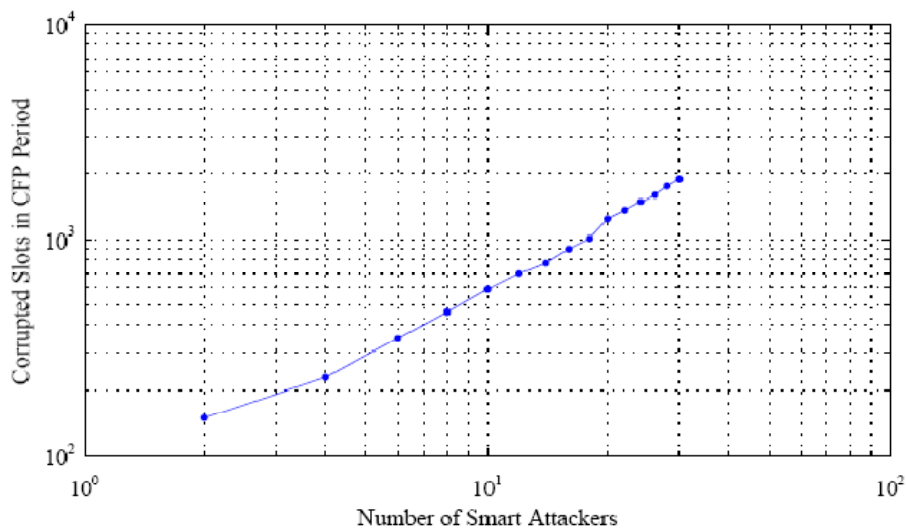

Figure 6 shows the total number of corrupted slots in the CFP period for a number of smart attackers. The figure shows that two smart attackers can successfully corrupt up to 149 GTS slots. This trend increases up to 1912 GTS slots for 30 smart attackers. Once the GTS slots are identified and attacked, the attackers try to decrease the bandwidth utilization in each slot. Corrupting more GTS slots result in the lowest bandwidth utilization. This corruption depends on the type of attacks. A smart attacker can corrupt more slots than a random or weak attacker. This is shown in Figure 7, where two smart attackers corrupt more slots and therefore decrease the bandwidth utilization by 71%. These are the best results in the attacker's point of view (and worst for the legitimate nodes). Two random attackers and one weak attacker decrease the bandwidth utilization by 49% and 15%, respectively. The later is the worst case for the attackers. As IEEE 802.15.4 networks may not frequently utilize the CFP period, the GTS attacks are not a big threat to them. But the direct adaptation of IEEE 802.15.4 security framework for WBAN is not reliable as most of the data is carried in the CFP period of the superframe.

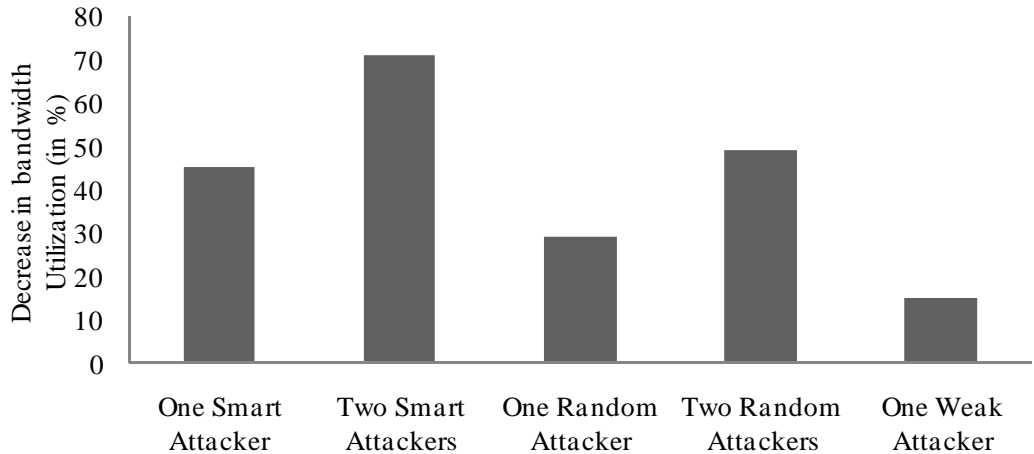

**Figure 7.** Decrease in bandwidth utilization.

## 7. Conclusions

Starting from the WBAN security requirements at different layers, we studied the IEEE 802.15.4 security framework for WBANs and identified different types of attacks on the IEEE 802.15.4 superframe by a number of adversaries. These attacks were categorized into smart, random, and weak attacks. Simulation results showed that the smart attacker(s) has the capability of corrupting an increasing number of GTS slots compared to random and weak attackers. This means that the direct adaption of IEEE 802.15.4 security framework for WBANs is not reliable since most of the traffic in WBANs is carried in CFP period, which is most vulnerable to GTS attacks.

One of the solutions is to implement a sophisticated backoff detection scheme that should successfully detect the backoff attacks. However, the backoff detection scheme may not work for adversaries who have enough knowledge of the scheme. They may try to maximize their throughput and minimize their chances of detection. Another approach is to allow the receiver to assign the backoff window to the sender. In this scheme, the receiver can easily detect any attack and can even penalize the adversaries by increasing their backoff values. A game theoretic approach could also be useful to detect and prevent the attacks by considering that all nodes are selfish.

### Acknowledgements

This work was supported by the National Research Foundation of Korea (NRF) grant funded by the Korea government (MEST) (No. No.2010-0018116) and by the Ministry of Knowledge Economy (MKE), Korea, under the Information Technology

Research Center (ITRC) support program supervised by the Institute for Information Technology Advancement (IITA) (IITA-2009-C1090-0902-0019).

## References

1. Ullah, S.; Higgins, H.; Braem, B.; Latre, B.; Blondia, C.; Moerman, I.; Saleem, S.; Rahman, Z.; Kwak, K.S. A comprehensive survey of wireless body area networks: On PHY, MAC, and Network Layers Solutions. *J. Med. Syst.* **2010**, doi: 10.1007/s10916-010-9571-3.
2. Ullah, S.; Higgins, H.; Shen, B.; Kwak, K.S. On the implant communication and MAC protocols for WBAN, *Int. J. Com. Syst.* **2010**, *23*, 982-999.
3. Saleem, S.; Ullah, S.; Yoo, H.S. On the security issues in wireless body area networks. *J. Digital Content Technol. Appl.* **2009**, *3*, 178-184.
4. *IEEE Standard 802.15.4: Wireless Medium Access Control (MAC) and Physical Layer (PHY) Specifications for Low Data Rate Wireless Personal Area Networks (WPAN)*; IEEE: Piscataway, NJ, USA, 2006.
5. Sastry, N.; Wagner, D.; Security considerations for IEEE 802.15.4 networks. In *Proceedings of the 3rd ACM workshop on Wireless security* (WiSe '04), Philadelphia (U.S.A.), Oct. 2004.
6. Shi, E.; Perrig, A. Designing secure sensor networks. *IEEE Wirel. Commun. Mag.* **2004**, *11*, 38-43.
7. Wood, A.D.; Stankovic, J.A. Denial of service in sensor networks. *IEEE Comput.* **2002**, *35*, 54-62.
8. Wang, Y; Attebury, G.; Ramamurthy, B. A survey of security issues in wireless sensor networks. *IEEE Commun. Surv. Tutorials* **2006**, *8*, 2-23.
9. Akyildiz, I.F.; Su, W.; Sankarasubramaniam, Y.; Cayirci, E. A survey on sensor networks. *IEEE Commun. Mag.* **2002**, *40*, 102-114.
10. Wang, X.; Gu, W.; Schosek, K.; Chellappan, S.; Xuan, D. *Sensor Network Configuration under Physical Attacks*; Technical report (OSU-CISRC-7/04-TR45); Department of Computer Science and Engineering, Ohio State University: OH, USA, July 2004.
11. Karlof, C.; Wagner, D. Secure routing in wireless sensor networks: Attacks and countermeasures. In *Proceedings of the 1st IEEE International Workshop on Sensor Network Protocols and Applications*, Alaska, May 2003; pp. 113-127.
12. Douceur, J. The Sybil attack. In *Proceedings of the 1st International Workshop on Peer-to-Peer Systems (IPTPS'02)*, Cambridge, February 2002.
13. Xiao, Y.; Chen, H.H.; Sun, B.; Wang, R.; Sethi, S. MAC security and security overhead analysis in the IEEE 802.15.4 Wireless Sensor Networks. *EURASIP J. WCN* **2006**, doi:10.1155/WCN/2006/93830.
14. Rijmen, V.; Daemen, J. The block cipher Rijndael. In *Smart Card Research and Applications*; LNCS 1820;, Springer-Verlag: New York, NY, USA, 2000; pp. 288-296.

15. Bellare, M.; Kilian, J.; Rogaway, P. The security of the cipher block chaining message authentication code. *J. Comput. Syst. Sci.* **2000**, *61*, 362-399.
16. Ha, J.; Kim, T.; Park, H.; Choi, S.; Kwon, W. An enhanced CSMA-CA algorithm for IEEE 802.15.4 LR-WPANs. *IEEE Commun. Lett.* **2007**, *Vol. 11*, No. 5, 461-463.
17. Huang, Y.; Pang, A.; Kuo, T. AGA: Adaptive GTS allocation with low latency and fairness considerations for IEEE 802.15.4. In *Proceedings of 2006 IEEE International Conference on Communications (ICC 06)*, Istanbul, Turkey, June 2006; pp. 3929-3934.
18. Jeon, J.; Lee, J.; Ha, J.Y.; Kwon, W.H. DCA: Duty-Cycle adaptation algorithm for IEEE 802.15.4 Beacon-Enabled Networks. In *Proceedings of VTC2007-Spring*, Dublin, UK, 22–25 April 2007; pp. 110-113.
19. Radosavac, S.; Cardenas, A.A.; Baras, J.S.; Moustakides, G.V. Detecting IEEE 802.11 MAC layer misbehavior in ad hoc networks: Robust strategies against individual and colluding attackers. *J. Comput. Secur.* **2007**, *15*, 103-128.
20. Bellardo, J.; Savage, S. IEEE 802.11 denial-of-service attacks: Real vulnerabilities and practical solutions. In *Proceedings of the USENIX Security Symposium*, Washington, DC, USA, August 2003.
21. Raya, M.; Hubaux, J.P.; Aad, I.; DOMINO: A system to detect greedy behavior in IEEE 802.11 hotspots. In *Proceedings of the Second International Conference on Mobile Systems, Applications and Services (MobiSys2004)*, Boston, MA, USA, June 2004.
22. Radosavac, S.; Baras, J.S.; Koutsopoulos, I. A framework forMAC protocol misbehavior detection in wireless networks. In *Proceedings of the 4th ACM Workshop on Wireless Security*, Cologne, Germany, August 2005; pp. 33-42.
23. Kyasanur, P.; Vaidya, N.; Detection and handling of mac layer misbehavior in wireless networks, In *Proceedings of the International Conference on Dependable Systems and Networks*, San Francisco, CA, USA, June 2003.
24. Sokullu, R.; Dagdeviren, O.; Korkmaz, I. On the IEEE 802.15.4 MAC layer attacks: GTS attack. In *Proceedings of Sensor Technologies and Applications, 2008. SENSORCOMM '08*, Cap Esterel, France, 25–31 August 2008; pp.673-678.
25. Ullah, S.; Shen, B.; Riazul Islam, S.; Khan, P.; Saleem, S.; Kwak, K. A study of MAC protocols for WBANs. *Sensors*, **2010**, *10*, 128-145.
26. *Network Simulator 2*; Available online: http://www.isi.edu/nsnam/ns/ (accessed on 20 September 2010).